\documentclass[prc,aps,preprint,showpacs]{revtex4}
\usepackage{graphicx}

\newcommand{\simleq}{\; \raisebox{-0.4ex}{\tiny$\stackrel
{{\textstyle<}}{\sim}$}\;}
\begin{document}
%\baselineskip 25pt minus .1pt
%\begin{center}

\title{Shape and shell-structure of lighter (N $\simleq$ 90) neutron-rich 
nuclei based on phenomenological Woods-Saxon potential} 

\author{ Ikuko Hamamoto$^{1,2}$ }

\affiliation{
$^{1}$ {\it Riken Nishina Center, Wako, Saitama 351-0198, Japan } \\ 
$^{2}$ {\it Division of Mathematical Physics, Lund Institute of Technology 
at the University of Lund, Lund, Sweden} }   

%\end{center}

%\vspace{2cm}

%\date{\today}

\begin{abstract}
By using a phenomenologically successful Woods-Saxon potential, 
I study the shape and shell structure of (A) neutron drip line nuclei 
with 10$\leq$N$\leq$60, (B) neutron-rich nuclei related to the r-process with 
40$\leq$N$\leq$90, and (C) one-particle spectra in the potential provided 
by the nucleus $^{70}$Fe as a representative of 
so-called N=40 ''island of inversion (IoI)'' nuclei.  
First, the shell structure that is unique in very weakly-bound neutrons is 
systematically studied, and the approximate neutron number of odd-N nuclei  
at which spherical (or deformed) halos can be found is pinned down.  
Second, the difference of the shell structure in r-process nuclei 
from that in stable nuclei is examined.  
Third, the similarity and the difference between the shell structure of 
N=20 IoI nuclei and that of N=40 IoI nuclei are analyzed.  As a result of it, 
it is concluded that in Fe and Cr isotopes  
the deformation called 
''N=40 IoI'' continuing up to N=50 is unlikely.

\end{abstract}

%\pacs{21.10.-k, 21.60.Ev, 21.10.Pc, 21.10.Gv}

\maketitle

\newpage

\section{INTRODUCTION} 

In the study of nuclear many-body problems, various elaborate microscopic 
models starting basically with some phenomenological 
two-body interactions (and sometimes together with some three-body 
interactions) have been available.  The parameters in 
such phenomenological models, in particular those of the two-body interactions, 
are usually determined so as to reproduce observed properties of nuclei around 
the stability line.  
When those models are numerically applied to nuclei far away from the 
stability line, it is often experienced that different models or parameters, 
which produce 
almost the same properties of stable nuclei, give quite different results 
of given nuclei away from the stability line.   A simple well-known example 
is Hartree-Fock calculations using various Skyrme interactions. A model which
can predict something that turns out to be in agreement with experimental data
is expected to contain some essentially correct element, 
compared to models having 
various adjustable parameters so as to reproduce available data.   
On the other hand, microscopic 
models such as shell models which contain always revisable parameters are also
useful for producing detailed wave functions which may be used in the analysis 
of data.
   
In the present work I would try to keep the model as simple as possible, 
when I study the shape and the shell-structure of very neutron-rich nuclei away
from the stability line.    
Then, since both shape and shell-structure are essentially one-particle 
properties of nuclei, I would work with the phenomenological one-body 
potential, of which 
the effect of neutron excess on nuclei around the stability line 
is in agreement with the observed data.  As far as nuclear shape is concerned, 
the study of one-particle shell structure is definitely the first important 
starting point.  
In other words, in the present work I explore  
how much can be said about 
nuclear shape and shell structure in stable and neutron-rich nuclei using  
the successful 
phenomenological one-body potential which includes the effect of the 
lowest-order neutron excess, instead of starting with phenomenological two-body 
interactions.     
 
I take from 
Ref. \cite{BM69} a spherically-symmetric Woods-Saxon potential in which the lowest-order effect of neutron excess, 
which is proportional to (N$-$Z)/A, is taken into account in the depth of 
the potential.  Here the neutron (proton) number is expressed by N (Z), 
while 
the mass number is denoted by A.  The coefficient of the (N$-$Z)/A term 
was determined phenomenologically 
from the observed properties of neutron-rich nuclei around the stability line.  
One of the simple features of the Woods-Saxon potential is that the depth of 
the nuclear potential for neutrons (and protons) is a function of the ratio, 
N/Z.  Namely, the depth is independent of the mass A for a given N/Z.  
Indeed, my experience of using the Woods-Saxon potential  
in the study of lighter neutron-rich or neutron-halo nuclei away 
from the stability 
line has shown that the Woods-Saxon potential works remarkably well 
with the result being in pretty good agreement with available experimental 
data.  For example, see \cite{IH07, IH10, IH16}.   

When deformation is suggested for certain neutron numbers by examining 
the shell structure for spherical shape, mainly due to the approximate 
degeneracy 
of one-particle eigenenergies 
of spherically symmetric potentials (Jahn-Teller effect), the shell structure 
in the axially symmetric quadrupole-deformed potential is further studied for 
those neutron numbers.     
The successful example of finding the region of nuclei with possible quadrupole 
deformation in the present way is the deformed stable rare-earth nuclei 
\cite{BM75}.   Namely, for some proton numbers which accept prolate  
deformation, nuclei  with 90$\leq$N$\leq$112 are deformed.  For those 
neutron numbers the energies of most one-particle levels occupied in the 
last major shell decrease 
(downward-going levels) for the relevant  
deformation parameter $\beta > 0$.   See, for example, Fig. 5-3 in Ref. 
\cite{BM75}.   This means that a very rough estimate of the total energy, 
which is just the sum of one-particle energies of occupied orbits, decreases 
for the relevant range of prolate deformation of those nuclei compared with 
spherical shape.   
This rare-earth example describes the 
deformed nuclei within one major shell, 82$\leq$N$\leq$126, where the 
pair correlation is important.   Due to the presence of pair correlation, 
deformed shape is not obtained immediately after neutrons start to occupy the
major shell.   The same idea was applied to lighter nuclei, 
neutron-rich Mg-isotopes with a smaller degeneracy 
(only the 1f$_{7/2}$ and 2p$_{3/2}$ shells), and the deformation 
of Mg-isotopes with N=21-26 \cite{IH07} was successfully predicted.  
Noting that the  
observed absolute dominance of 
prolate deformation over oblate deformation in the ground state of 
even-even deformed nuclei is not yet really understood \cite{IH09}, in the
present work I do not consider the possibility of oblate shape for the ground
state of even-even nuclei. 

One-particle energy spectra in the quadrupole-deformed potential are known 
to have a close relation to the observed low-lying energy spectra in odd-A 
deformed nuclei \cite{BM75}.  
In other words, the one-particle picture in deformed nuclei seems to work 
much better 
than that in spherical nuclei.  Namely, in nuclei with 
spherical shape one-particle energies may be compared with low-lying 
spectra at best only in the (spherical closed-shell plus/minus one nucleon) 
nuclei.  
In contrast, in deformed nuclei it is known that one-particle picture works 
well in most odd-A well-deformed stable rare-earth nuclei \cite{BM75}.   
This is certainly because in the 
phenomenological quadrupole-deformed potential the major part 
of the residual interaction in the spherical potential, 
quadrupole-quadrupole 
interaction, is already included in the one-body potential.     

In the present article first I show the neutron shell structure 
in the spherically symmetric potential unique in neutron drip line nuclei 
with 10 $<$ N $<$ 60 by taking N/Z = 2.   The observed heaviest odd-N nuclei 
for given even-Z are, for example, $^{19}_{6}$C$_{13}$, $^{23}_{8}$O$_{15}$, 
$^{31}_{10}$Ne$_{21}$,  $^{37}_{12}$Mg$_{25}$ and so on.  Therefore, the ratio 
N/Z = 2 is an approximate estimate of the even-even core for odd-N nuclei 
with the last bound odd  neutron.  
The ratio may, of course, change in heavier isotopes.   
Second, I am interested in the neutron shell structure of a  
spherically-symmetric potential in nuclei around the r-process.  
Here the Fermi level may be expected around $-$3 to $-$5 MeV, which corresponds to 
the ratio N/Z $\approx$ 1.7 in the lighter mass region of the r-process.    
Since actual nuclei involved in the r-process have a finite temperature, 
the information on the shell structure of particle spectra may not be 
directly obtained from 
observation.   Nevertheless, the information on the shape and shell structure 
in the considered neutron-rich nuclei is basically important for understanding  the structure of the r-process.   The shell structure of the neutron  
spectra obtained will be compared with that of stable nuclei in the region 
of the same neutron number.   
Third, as an example of going one step further from the shell structure 
related to the   
spherically symmetric potential, taking the shell structure obtained for a  
quadrupole 
deformed potential I study the nuclei $^{70}_{26}$Fe$_{44}$ and 
$^{66}_{24}$Cr$_{42}$,  
for which 
the ratio is N/Z $\approx$ 1.7, though those nuclei are a bit too light to 
have something to do with the r-process.  
Those nuclei were recently studied intensively both experimentally
\cite{GB15, CS15, GS16} and 
theoretically and have a typical shell structure which indicates 
the possibility of 
having deformed shape similar to the shell structure of nuclei with 
N$\approx$20 
called the "island of inversion (IoI)".       
My intention is to look for a general basic sign of deformation 
appearing already in one-particle spectra of the spherical  
one-body potential before going 
to complicated microscopic calculations which necessarily contain 
various parameters.                  

In Sec. II the main points of the model used are briefly summarized,   
while the results of numerical calculations are given in Sec. III.    
In Sec. III. A the shell-structure unique 
to the neutron drip line is exhibited  
for 10 $<$ N $<$ 60, while in Sec. III. B the shell-structure 
in neutron-rich nuclei with 40 $<$ N $<$ 90 expected in the region of 
lighter nuclei 
around the r-process is shown and discussed in comparison with that in 
stable nuclei.   
In Sec. III. C nuclei $^{70}_{26}$Fe$_{44}$ and $^{66}_{24}$Cr$_{42}$,  
which belong to the part of the 
lightest mass region in Sec. III. B, namely the region of "IoI 
around N=40", are chosen to discuss the possible spherical or deformed shape 
based on the shell structure.   Conclusion and discussions are given in 
Sec. IV.

\section{MODEL}
As a one-body nuclear potential I take the Woods-Saxon potential described 
in p.238-240 of Ref. \cite{BM69}, for which the neutron-excess term determined 
phenomenologically
by the properties of stable nuclei has worked pretty well in the analysis of 
light neutron drip line nuclei. For example, see Refs. \cite{IH07, IH10}.   
In order to systematically study the shape and shell structure of 
neutron-rich nuclei away from the stability line, in the present work I choose 
to start not with a two-body interaction but directly with the one-body 
potential, for which the parameters are 
phenomenologically determined.   
When observed properties of the ground state of nuclei indicate that nuclei 
are not spherical, the relevant deformation is almost always 
axially-symmetric quadrupole deformation.   Therefore, as deformation 
in the present article I consider only axially symmetric quadrupole deformation.
Though the Hamiltonian and the related formula are the same as those given 
in Ref. \cite{IH04}, for completeness I give a brief summary 
of them in the following.   
The axially symmetric quadrupole-deformed potential for neutrons consists of 
the following three parts.  
\begin{equation}
V(r) = V_{WS} \, f(r), 
\label{eq:VWS}
\end{equation}
\begin{equation}
V_{coupl}(\vec{r}) = -\beta k(r) Y_{20}(\hat{r}), 
\label{eq:vcoupl}
\end{equation}
\begin{equation}
V_{so}(r) = - V_{WS} \, v \, \left(\frac{\Lambda}{2} \right)^{2} \, 
\frac{1}{r} \, \frac{df(r)}{dr} \, 
(\vec{\sigma} \cdot \vec{\ell}),  
\end{equation}
where  $\Lambda$ is the reduced Compton wave-length of nucleon 
$\hbar$/$m_{r} c$, 
\begin{equation}
f(r) = \frac{1}{1+exp \left(\frac{r-R}{a} \right)}
\label{eq:fr}
\end{equation}
and 
\begin{equation}
k(r) = R V_{WS} \frac{df(r)}{dr} \, . 
\label{eq:kr}
\end{equation} 
Though the expression of the spin-orbit potential, (3), is slightly different 
from the one in Ref. \cite{BM69}, the value of $v=32$ which I use produces 
the same spin-orbit potential given in Ref. \cite{BM69}.   
When the shell structure of protons is studied, the Coulomb potential 
for protons is included in the one-body potential.   
The values of parameters in the one-body potentials are taken 
from Ref. \cite{BM69}.  The point related to the present work is 
that the parameters in 
the potential connected to neutron excess are taken from the properties of 
the stable neutron-rich nuclei.  

I write the single-particle wave-function in the body-fixed (intrinsic) 
coordinate system as 
\begin{equation}
\Psi_{\Omega}(\vec{r}) = \frac{1}{r} \, \sum_{\ell j} R_{\ell j \Omega}(r) \,  
{\bf Y}_{\ell j \Omega}(\hat{r}),
\label{eq:twf}
\end{equation}
which satisfies 
\begin{equation}
H \, \Psi_{\Omega} = \varepsilon_{\Omega} \, \Psi_{\Omega}   
\end{equation} 
where $\Omega$ expresses the component of one-particle angular-momentum 
along the symmetry axis and is a good quantum number.   
The coupled differential equations for the radial wave-functions 
are written as 
\begin{equation}
\left(\frac{d^2}{dr^2} - \frac{\ell (\ell +1)}{r^2} + \frac{2m}{\hbar^2}( 
\varepsilon_{\Omega} - V(r) - V_{so}(r) ) \right) R_{\ell j \Omega}(r) = 
\frac{2m}{\hbar^2} 
\sum_{\ell^{'} j^{'}} \langle {\bf Y}_{\ell j \Omega} \mid V_{coupl} 
\mid {\bf Y}_{\ell^{'} j^{'}
\Omega} \rangle R_{\ell^{'} j^{'} \Omega}(r)
\label{eq:cpl}
\end{equation}
where 
\begin{eqnarray}
\langle {\bf Y}_{\ell j \Omega} \mid V_{coupl} \mid {\bf Y}_{\ell^{'} j^{'} 
\Omega} 
\rangle 
 & = &
- \beta \, k(r) \, \langle {\bf Y}_{\ell j \Omega} \mid Y_{20}(\hat{r}) \mid 
{\bf Y}_{\ell^{'} j^{'} \Omega} \rangle \nonumber \\
& = & 
- \beta \, k(r) \, (-1)^{\Omega-1/2} \, 
\sqrt{\frac{(2j+1)(2j^{'}+1)}{20 \pi}} \nonumber \\
&& \mbox{ } 
C(j, j^{'}, 2;\Omega, -\Omega, 0) \, C(j, j^{'}, 2; \frac{1}{2}, -\frac{1}{2}, 
0).   
\end{eqnarray} 

The eigenvalue $\varepsilon_{\Omega}$($<$0) of the coupled equations 
(\ref{eq:cpl}) 
for a given value of $\Omega$ is obtained by integrating the equations 
in coordinate space  
for given values of 
$\beta$ and radius $R$, with the asymptotic behavior 
of $R_{\ell j \Omega} (r)$ for 
$r \rightarrow \infty$ 
\begin{equation}
R_{\ell j \Omega} \propto r \, h_{\ell}(\alpha r)
\end{equation}
where $h_{\ell}(-iz) \equiv j_{\ell}(z) +i n_{\ell}(z)$, in which 
$j_{\ell}$ and $n_{\ell}$ are spherical Bessel 
and Neumann functions, respectively, 
and 
\begin{equation}
\alpha^{2} \equiv -\frac{2m \, \varepsilon_{\Omega}}{\hbar^{2}}
\end{equation}
The normalization condition is written as 
\begin{equation}
\sum_{\ell, j} \, \int^{\infty}_{0} \, \mid R_{\ell j \Omega}(r) \mid ^{2} \, 
dr =1
\end{equation}

For spherical shape ($\beta = 0$) the eigenvalues depend only on 
($\ell, j$) and the Schr\"{o}dinger equation is integrated  
in the laboratory coordinate system with the correct asymptotic behavior 
of wave functions with $(\ell, j)$ at  
$r \rightarrow \infty$.   For axially symmetric quadrupole-deformed shape 
an infinite number of channels with ($\ell, j,\Omega$) are coupled for a given 
$\Omega$.  
In practice, the number of coupled channels or the maximum value of ($\ell, j$) 
pairs 
included for a given $\Omega$ is determined so that the resulting eigenvalues 
and wave functions do not change in a meaningful way 
by taking a larger number of coupled channels.  
The coupled differential equations obtained from the 
Schr\"{o}dinger equation are integrated in coordinate space with the correct 
asymptotic behavior of radial wave functions in respective ($\ell, j$) 
channels for $r \rightarrow \infty$.  The solution obtained in this way is 
independent of the upper limit of radial integration, $R_{max}$, 
if $f(r)$ in (\ref{eq:fr}) and $k(r)$ in (\ref{eq:kr}) are already negligible 
at $r = R_{max} \gg R$.      
The r-dependence of a given ($\ell, j$) 
component thus obtained is generally different from that of any eigenfunctions  
in the spherically symmetric potential.  

The Hamiltonian and how to solve the coupled differential equations 
derived from 
the Schr\"{o}dinger equation for one-particle resonant levels in the deformed 
potential are described in Ref. \cite{IH05}.  Since one-particle resonant 
levels are not the main subject of the present article, 
for the details I just refer  
to Ref. \cite{IH05}, in which one-particle resonant energy
$\varepsilon_{\Omega}{^{res}}$ is sought so that 
eigenphase $\delta_{\Omega}$ 
increases through ${\pi}$/2 as $\varepsilon_{\Omega}$ increases. 

I take the diffuseness $a = 0.67$ fm and the radius $R = r_{0}A^{1/3}$ 
with $r_{0} =$ 1.27  fm, which are the standard parameters used in $\beta$ 
stable nuclei \cite{BM69}.
The depth of the Woods-Saxon potential in (\ref{eq:VWS}) is 
\begin{equation}
V_{WS} = -51 \pm 33 \, \frac{N-Z}{A}  \qquad \mbox{MeV}
\label{eq:WS}
\end{equation}
where the upper sign (+) is for neutrons while the lower one ($-$) is 
for protons. Since A=N+Z, the depth is a function of N/Z independent of 
the mass number A.  

Though the shape of a nucleus is determined by  
both neutrons and protons, here I examine the dependence of 
shape on the shell structure of neutrons and protons, separately.       
In the discussion of the possible shape of a given nucleus I use the following 
empirical facts.  First, from the shell structure in spherically symmetric 
potential, (a) the presence of a considerable energy gap in one-particle energy 
spectra for spherical shape indicates that spherical shape is preferred 
for the nucleon number filling particles up to the levels below the gap.  
This fact is 
well known from the energy spectra for  nuclei with magic numbers of 
nucleons.  (b) The presence of some particles 
in a few almost degenerate ($\ell, j$) shells around the Fermi level suggests 
that the nuclei may be deformed, as those particles have the possibility of 
gaining energy by breaking spherical symmetry (Jahn-Teller effect).  This fact 
is known, for example, 
from the observation of deformed nuclei when both neutron- and 
proton-numbers are away from magic numbers by several particles.  
If pair correlation is 
not effective in some nuclei, it may be possible for nuclei with one 
or two nucleons in the almost degenerate ($\ell, j$) shells to be deformed.  
This phenomena includes the case that one-particle levels of 
some almost degenerate 
($\ell, j$) shells lie in the continuum as one-particle resonant levels, 
if energies of 
the one-particle levels  
connected to the degenerate ($\ell, j$) shells for spherical shape 
come down to be bound when 
the potential is deformed.   A good example of this case is the deformation 
of $^{34}_{12}$Mg$_{22}$ shown in Fig. 2 of Ref. \cite{IH12}. 
Second, from the shell structure in axially symmetric quadrupole-deformed 
potential, (c) the presence of a considerable energy gap in one-particle 
energy spectra for a given deformation indicates that the configuration 
filling particles in the energy levels 
up to the levels below the gap is relatively stable 
for the nucleon number at the deformation; (d) The deformation, 
at which some energy levels are almost degenerate, may not be preferred 
by the nucleon number corresponding to partial filling  of 
those energy levels, because those nucleons have the possibility of 
gaining energy 
by breaking the symmetry of  axially symmetric quadrupole deformation 
(Jahn-Teller effect); 
(e) A very approximate measure of finding deformed nuclei is that the energies 
of most one-particle 
levels in the last filled major shell 
are going down as $|\beta|$ increases from zero.   This measure is 
in good agreement with finding the prolately-deformed region of stable 
rare-earth nuclei.  See, for example, Fig. 5-3 of Ref. \cite{BM75}.  
However, this approximate measure should not be used 
for very large deformations, at which various other factors 
such as surface tension, 
higher-order deformation etc.  may start to play an appreciable role.

\section{NUMERICAL RESULTS}

\subsection{Light neutron-drip-line nuclei (N/Z = 2) with 10 $<$ N $<$ 60}
As is already written in Introduction, an approximate ratio N/Z of 
the even-even core of the observed heaviest 
odd-N nuclei of $_{6}$C, $_{8}$O, $_{10}$Ne, and $_{12}$Mg isotopes 
is about 2.   Thus, in order to see the systematic shell-structure of neutrons 
in neutron drip line nuclei, in Fig. 1 I show the neutron one-particle 
energies as a function of the neutron number, taking N/Z = 2.  
For N/Z = 2 the depth of the Woods-Saxon potential V$_{WS}$ is $-$40 MeV.   
In Fig. 1 the Fermi levels of nuclei with respective N values lie around 
$\varepsilon_{j} = 0$.  This can be seen by checking the Fermi level 
by filling  N neutrons from the bottom, 1s$_{1/2}$ level, though 
the 1s$_{1/2}$ energy is outside the figure,  for example,  
$-$33.1 MeV for N = 60 and $-$21.1  MeV for N = 10, respectively.         
For example, the one-particle level scheme for the $_{12}$Mg isotope is 
the one for 
the potential provided by  
$^{36}_{12}$Mg$_{24}$, while that for the $_{26}$Fe isotope is the one 
for the potential provided by $^{78}_{26}$Fe$_{52}$.  
 
It is seen that in the spherical potential produced by $^{12}_{4}$Be$_{8}$ 
the energies of 2s$_{1/2}$ and 1d$_{5/2}$ levels are almost degenerate, 
in the spherical potential by $^{36}_{12}$Mg$_{24}$   
the energies of 2p$_{3/2}$ and 1f$_{7/2}$ levels are almost degenerate, 
and in the potential produced by $^{78}_{26}$Fe$_{52}$ the energies 
of 3s$_{1/2}$ and 2d$_{5/2}$ levels are almost degenerate.  
The degeneracy of respective two levels in the spherical potential 
was expected 
for respective neutron drip line nuclei, and the possible resulting 
deformation and the related halo phenomena have been partly discussed in Refs. 
\cite{IH07, IH12}.  Nevertheless, 
it is useful to see them in this more systematic way, 
even if the neutron numbers of neutron drip line nuclei for respective 
isotopes may not strictly agree with experimental information.  
The prominent change of the neutron level structure in the spherical potential
especially for $\mid \varepsilon_{j} \mid$ $\simleq$ 7 MeV and 
the related physics are
described in Fig. 2-30 and p.239-240 of Ref. \cite{BM69}.  A figure
similar to a part of Fig. 1 is found in Ref. \cite{AO00} in relation to the
description of a new magic number N=16. 

One sees that the behavior of energy eigenvalues of weakly-bound 
neutron orbits with smaller orbital angular-momentum is different 
from that with larger orbital angular-momentum, 
for example, by examining the energy variation of the 2s$_{1/2}$ 
level relative to that of 1d levels as a function of N in Fig. 1.  
For the system with N = 60 the 2s$_{1/2}$ level lies above both the 1d$_{5/2}$ 
and 1d$_{3/2}$ levels.  In contrast, for the system with N = 10 
the 2s$_{1/2}$ level lies lower than both 1d levels.    
Though by going from N = 60  to N = 10 in Fig. 1 the depth of the Woods-Saxon 
potentials remains the same, the volume of the potential becomes smaller 
by about a factor of 6 since the mass number changes from 90 to 15.   
The energies 
of both 1d orbits change more or less smoothly from $-$20 MeV to 0 MeV.  
In contrast, the energy curve of the 2s$_{1/2}$ orbit changes strongly,  
in particular for $\mid \varepsilon_{j} \mid \simleq 3$ MeV.  
This is because since 
there is 
no centrifugal barrier 
for s-orbits, the probability for s-neutrons to remain inside the potential is
zero in the limit that eigenenergies $\varepsilon_{s1/2}$ ($<$0) approach zero. 
Namely, weakly-bound neutrons in the s-orbit spend most of the time 
outside the potential produced by core nuclei as the one-particle energy 
$\varepsilon_{j} < 0$ approaches zero.   
The slope of the curve of s$_{1/2}$ orbits in Fig. 1, 
$d\varepsilon_{j}/dN$, approaches zero 
in the limit of $\varepsilon_{j} (<0) \rightarrow 0$.   
In contrast, neutrons in the d-orbits 
lose the binding energy relatively more rapidly as the binding field becomes 
weaker, because due to the presence of the centrifugal barrier the major part 
of the wave function stays inside the potential \cite{BM69, IH07, IH16}.  
The same physics coming from the considerable difference 
between the heights of 
centrifugal barriers produces the phenomena that in Fig. 1 the 2p$_{3/2}$ 
energy approaches the 1f$_{7/2}$ energy as the one-particle binding energies 
approach zero.

\subsection{Neutron-rich nuclei (N/Z = 1.7) with 40 $<$ N $<$ 90}
A basic question is whether or not the shell-structure in neutron-rich 
nuclei around the r-process is approximately the same as that of stable 
nuclei if the shell structure around the same neutron number 
is examined in the two cases.  The shell structure around respective Fermi
levels is, in practice, of interest.     
I examine the question by using the present phenomenological one-body 
potential, since the study of the shape and shell-structure by using 
the present potential has been successful at least in light neutron drip line 
nuclei.   

In Fig. 2 the neutron one-particle energy eigenvalues as a function of 
neutron number are shown, taking N/Z = 1.7 which determines the depth 
of the potential to be  
V$_{WS}$ = $-$42.4 MeV.     The N/Z ratio is an approximate ratio 
in lighter nuclei related to the r-process just above 
$^{78}_{28}$Ni$_{50}$.   
The ratio N/Z = 1.7 in this mass region means that the neutron excess 
corresponds roughly to one major shell in the sense of the traditional 
shell model.   
The Fermi levels of relevant nuclei in Fig. 2 lie 
around $\varepsilon_{j} \approx -4$ MeV.  
For example, the one-particle energy spectrum at N=68 in Fig. 2 is the one   
produced by the potential of $^{108}_{40}$Zr$_{68}$.         

For comparison, in Fig. 3 the neutron one-particle energy eigenvalues of 
the potential produced by nuclei along the stability line are shown 
as a function of 
neutron number.   The Fermi levels of relevant nuclei lie 
around 
$\varepsilon_{j} \approx -8$ MeV.  Roughly speaking, deformation is not 
particularly favored by the proton 
numbers of nuclei corresponding to the neutron numbers in Fig. 3, 
in contrast to the fact that  
deformation may be favored by the proton numbers of nuclei in the region of
60$\simleq$N$\simleq$70 of Fig. 2.   
This is because, for example, in the region of 50$\simleq$N$\simleq$70 
of stable nuclei  
protons gradually fill in the single high-j (1g$_{9/2}$) shell.  
The comparison of 
eigenvalues in Fig. 3 with observed low-lying spectra of relatively 
simple spherical nuclei seems good.   
For example, the observed spin-parities of 
the ground, first excited, and second excited states of $^{121}_{50}$Sn$_{71}$ 
are 3/2$~{+}$,   
11/2$^{-}$, and 1/2$^{+}$, respectively, which should be compared with 
$\varepsilon_{j}$ values at N=70 in Fig. 3.  Another example is 
that the observed 
spin-parities of the ground and first excited states of $^{91}_{40}$Zr$_{51}$ 
are 5/2$^{+}$ and 1/2$^{+}$, respectively, which should be compared with 
$\varepsilon_{j}$ values at N=50 in Fig. 3.   

A very systematic behavior of $\varepsilon_{j}$ which is found 
in Figs. 1, 2 and 3 is that, as a function of neutron number,  
$\varepsilon_{j}$ ($<$0) of  
lower-$\ell$ orbits decreases more slowly than $\varepsilon_{j}$ of 
higher-$\ell$ orbits.   
In contrast, in the harmonic oscillator (HO) potential 
without a spin-orbit potential the levels of the 2s and 1d neutrons are 
degenerate.   
The higher energy of the 2s level than the 1d levels at N=60 seen 
in Fig. 1 is 
explained by considering 
the difference of the radial shape of the Woods-Saxon potential from that of 
the HO potential (see p.222 of Ref. \cite{BM69}).   
On the other hand, in order to find the effect of the presence of nuclear 
surface together with the quantum-mechanical leaking of bound 
one-particle wave-functions to the outside of the surface 
on the eigenvalues $\varepsilon_{j}$, 
I pay attention to the eigenvalues of neutrons 
in the infinite square-well potential (cavity) without spin-orbit potential.    
In the cavity there is no leaking of wave functions to the outside of 
the potential.  
The eigenvalues $\varepsilon_{n \ell}$ are obtained from 
the $n$th zero of the spherical Bessel function of order $\ell$.   Then, 
among the family with $\Delta \ell$ = 2 originating from a given HO major shell 
the one-particle energies decrease as $\ell$ increases \cite{IH09}.   
For example, the $(n+1)s$ level is energetically always higher than the 
$nd$ level.    
This simple example shows an important role of the nuclear surface 
in the shell structure of one-particle energies in finite potentials.   

When the shell structure for a given N around the Fermi level in Fig. 2 is 
compared with that in Fig. 3, the main systematic difference comes from the 
fact that the 
decrease of $\varepsilon_{j}$ of low-j orbits is more moderate than 
that of high-j orbits  
as neutron number increases.   
A typical example is that for N=65 in Fig. 2 the 1g$_{7/2}$ level is nearly 
degenerate to the 3s$_{1/2}$ level at $\varepsilon_{j} \approx -4$ MeV, while  
for N=65 in Fig. 3 the former is lower than the latter by more than 1 MeV 
around $\varepsilon_{j} = -8$ MeV.   This difference comes mainly from the 
difference of the Fermi energies in Figs. 2 and 3.   As the neutron number 
decreases (namely, 
as the potential gets weaker), the binding energies of all j orbits smoothly 
decrease, however, the higher-$\ell$ orbits decrease the binding energies 
appreciably more than the lower-$\ell$ orbits.   This kind of shell-structure 
change within a given one major shell is systematically present in finite-well 
potentials as described in the previous paragraph, though there can be, 
in practice,  
other locally-stronger shell-structure changes in neutron one-particle spectra, 
which come, for example, 
from a strong attractive neutron-proton interaction depending on 
the proton orbits 
occupied.   
The systematic change of shell-structure can have a serious effect, 
for example, 
on the spin-parities of respective low-lying states and the related observed
quantities, while the effect on  
some other observed quantities may be to some extent smoothed out when 
the system has a finite temperature as in the r-process.

\subsection{Cr- or Fe-isotopes with N/Z $\approx$ 1.7}
I focus on the shell structure in the potential provided by 
$^{66}_{24}$Cr$_{42}$ and 
$^{70}_{26}$Fe$_{44}$, for which the ratio N/Z is equal to 1.75 and 1.69, 
respectively.  
Since the calculated shell-structure of these two nuclei 
is very similar, in the following the numerical result of only 
$^{70}$Fe is presented.   
The recent experimental information \cite{GB15, CS15} indicates that even-even 
Cr isotopes with N=38-42 and even-even Fe isotopes with N=40-46 are deformed, 
based on both the observed very low energies of the first excited state 
which is assumed to be 2$^{+}$ in some cases  
and the observed ratios of E(4$_1^+$)/E(2$_1^+$) where 
the second excited state 
is assumed to be 4$^+$ in some cases.        
An interesting question is whether or not 
Cr isotopes with N$>$42 and Fe isotopes with N$>$46 are also deformed.  
The question is related to the basic origin of the deformation 
in terms of 
neutron shell structure, as the neutrons with N$>$40 
are expected to occupy 
the single-j-shell, 1g$_{9/2}$, in the simple picture of spherical shape. 
       
In Fig. 4 the calculated proton one-particle energies in the potential 
produced by $^{70}$Fe as a function of deformation $\beta$ are shown, 
while the calculated neutron one-particle energies in the potential provided 
by $^{70}$Fe are exhibited in Fig. 5.     
It is seen from Fig. 4 that some moderate-size prolate deformation may be 
favored by proton numbers Z=24 (Cr) and 26 (Fe) while some oblate deformation 
may be certainly disfavored.  The preference of prolate deformation is 
consistent also with the recent experimental finding in Ref. \cite{XYL18} of 
the rotational spectra based on the 5/2$^{-}$ ground state 
in $^{65,67}_{25}$Mn$_{40,42}$.  
In Fig. 4 it is also noted that the proton one-particle 
energy difference between the 2d$_{5/2}$ and 1g$_{9/2}$ levels is as large as 
5.37 MeV, while the energy difference between the 1g$_{9/2}$ and and 2p$_{1/2}$ 
levels is only 1.57 MeV.  In short, Z=50 is a good magic number as in stable 
nuclei.    

From Fig. 5 it is seen that the neuron one-particle energy difference between 
the 2d$_{5/2}$ and 1g$_{9/2}$ levels is 2.62 MeV, while the one between 
the 1g$_{9/2}$ and 2p$_{1/2}$ levels is 3.15 MeV.   
Namely, for spherical shape the N=50 energy gap is smaller than the N=40 energy 
gap.  Since the one-particle levels above N=50 such as the 2d$_{5/2}$  
and 3s$_{1/2}$ levels have 
the same positive parity as 1g$_{9/2}$, there is a possibility of being mixed 
with 1g$_{9/2}$ by the quadrupole-deformed field so as to lower the energies 
of lowest-lying levels with given $\Omega$ values, which are the levels 
connected to the 1g$_{9/2}$ state at $\beta$=0.   
The appreciable amount of such mixture in the $\Omega^{\pi}$ = 1/2$^{+}$, 
3/2$^{+}$, and 5/2$^{+}$ orbits can be seen from the dependence of 
those one-particle energies on $\beta$ in Fig. 5.   
Due to the relatively large energy 
distance between the neutron 1g$_{9/2}$ and 2p$_{1/2}$ levels and the strong 
quadrupole coupling between the 1g$_{9/2}$ and 2d$_{5/2}$ orbits, 
the possibility of lowering the energy of the system in the region 
of N$\simleq$46 is seen by filling  
one-particle levels for the moderate-size prolate deformation.  In short,   
examining the behavior of one-particle energies for prolate deformation 
$\beta$ $\approx$ 0.2, one may expect that both Cr- and Fe-isotopes 
can be prolately deformed up to N$\approx$46.    

Experimental information on the spin-parity of the ground state of 
neutron-rich (N$>$40) odd-N Fe isotopes is not yet sufficiently established.   
However, some 
experimentally suggested spin-parities, 5/2$^{+}$ for $^{67}_{26}$Fe$_{41}$ 
\cite{HJ05}  
and 1/2$^{-}$ for $^{69}_{26}$Fe$_{43}$ \cite{CDN14}, 
are difficult to obtain for spherical 
shape.   In contrast, as is seen in Fig. 5 it is pretty easy to expect 
those spin-parities if those 
nuclei are prolately deformed by a moderate amount.   
This expectation is valid also in the case that a 
proper pair correlation is included in the consideration.   

I compare the shell structure of neutron-rich nuclei in the present N=40 region 
with that of the N=20 IoI region.  
The similarities of the neutron shell structure for spherical shape 
of neutron-rich nuclei
in the two regions are, (a) the lower one-particle shells are high-j shells,  
1f$_{7/2}$ and 1g$_{9/2}$, which are energetically pushed down 
from respective major shells by the large $\ell$-s splitting in the sense 
of Ref. \cite{JM50};    
(b) The higher one-particle shells, 2p$_{3/2}$ and 2d$_{5/2}$, are weakly bound
(or resonant) and have smaller orbital angular-momenta than 
the respective lower
ones by $\Delta \ell = 2$.  Thus, the energy distance between the higher 
and lower shells becomes smaller compared with that in deeply-bound cases 
\cite{IH07}.  
Furthermore, the higher shells can be strongly quadrupole-coupled 
with respective lower
ones due to $\Delta \ell = 2$ and $\Delta j = 2$.  On the other hand,  
the differences are (c) 
due to the smaller $\ell$-s splitting in the f-shell than in the g-shell 
the lowering of the 1f$_{7/2}$ level is smaller than that of the 1g$_{9/2}$ 
level; (d) for a given small binding energy the degree of energy lowering 
of the 2p$_{3/2}$ 
($\ell = 1$) level due to the leaking of the wave function 
to the outside of the surface 
is larger than that of the 2d$_{5/2}$ ($\ell = 2$) level.   

As an example of the neutron potential provided by deformed nuclei in the N=20
IoI region I take the potential given by $^{34}_{12}$Mg$_{22}$ (where 
N/Z = 1.83), which is shown in Fig. 2 of Ref. \cite{IH12}.   
For spherical shape ($\beta = 0$) the large energy gap (4.51 MeV) at N=20 
is kept well, while the energy difference between the 2p$_{3/2}$ and 1f$_{7/2}$ 
is only 0.391 MeV.   Namely, the 2p$_{3/2}$ and 1f$_{7/2}$ levels are almost
degenerate and the spherical energy gap at N=28 or the magic number N=28 has totally 
disappeared. 
Consequently, neutrons in those neutron-rich Mg isotopes partially filling 
the combined (2p$_{3/2}$ + 1f$_{7/2}$)
shell may easily make the energy of the system lower by making deformation.  
A simple approximate way often used to guess possible deformation using the figures such as Fig. 2 of Ref. \cite{IH12} is to examine the deformation-dependence of the energy of the last-filled (Fermi) one-particle-level for a given N.  
I note that, for example, in the degenerate harmonic-oscillator model the first (second) half of a given major shell has prolate (oblate) shape, and the slope of one-particle energy as a function of prolate deformation changes from downward to upward just in the middle of the major shell.    
Considering only prolate shape the Fermi energy of the deformed system with 
20$\leq$N$\leq$28 can become lower than that of spherical shape, which is equal to $\varepsilon$(1d$_{3/2}$) for N=20 and $\varepsilon$(1f$_{7/2}$) for 
21$\leq$N$\leq$28.      
Since the (2p$_{3/2}$ + 
1f$_{7/2}$) shell accommodates 12 neutrons, the isotopes with up to 
6-8 neutrons in the combined shell could be expected to be prolately deformed 
\cite{IH07}, 
neglecting pair correlation in these light nuclei.  Later this expectation
was experimentally found to be fulfilled \cite{PD13}.   
In contrast, at $\beta = 0$ in Fig. 5 the energy gap (2.62 MeV) at N=50 
is small but it is not close to zero. 
This appreciable energy gap leads to the fact that the Fermi energy of 
the N=50 system within moderate prolate deformation  
in Fig. 5 never becomes lower than that of spherical shape, 
which is equal to $\varepsilon$(1g$_{9/2}$).   
If the 2d$_{5/2}$ and 1g$_{9/2}$ shells had been nearly degenerate, 
for $\beta > 0$ all downward-going levels originating from the two shells 
would have been gathered in the energetically lower half 
as in the case of the level 
structure coming
from the combined (2p$_{3/2}$ + 1f$_{7/2}$) shell seen in Fig. 2 
of Ref. \cite{IH12}.   
Then, deformed shape might have been preferred for up to 10 neutrons 
in the combined (2d$_{5/2}$ + 1g$_{9/2}$) shell, 
which is slightly more than a half of 
the neutron number that can be accommodated 
in the combined (2d$_{5/2}$ + 1g$_{9/2}$) 
shell. Namely, deformed shape might have been possible continuously   
up to N=50.  However, in the present case it is expected that the Fe and Cr 
isotopes
only up to N$\approx$46 may be prolately deformed by an appreciable amount.   

The characteristic features of calculated neutron one-particle energies in the  
potential provided by $^{76}_{26}$Fe$_{50}$ are 
very similar to those of Fig. 5. 
For example, the depth of the potential is $-$40.6 MeV and one-particle 
energies at $\beta = 0$ are $-$3.11, $-$0.47, and $-$0.29 MeV for the 
1g$_{9/2}$, 2d$_{5/2}$, and 3s$_{1/2}$ levels, respectively.   Namely, 
the energy distance between the 2d$_{5/2}$ and 1g$_{9/2}$ levels is 2.64 MeV.  
In Ref. \cite{CS15} it was wondered whether or not the deformation extends to
N=50 in the $_{24}$Cr and $_{26}$Fe isotopes, though in Ref. \cite{CS15} 
no results of 
shell-model calculations were given for those isotopes, for which experimental
data have not been obtained.   In my present work I conclude that the
deformation continuously up to N=50 is unlikely.

 \section{CONCLUSION AND DISCUSSIONS}
In the present work I have studied the shape and shell structure of lighter 
neutron-rich nuclei away from the stability line, 
by using the Woods-Saxon potential, 
for which parameters are phenomenologically 
determined so as to reproduce the properties of neutron-rich stable nuclei.   
The success and usefulness of the Woods-Saxon potential have been already 
known in the study of the properties related to one-particle operators in 
light neutron drip line nuclei.   

First, I showed calculated neutron one-particle energies 
in lighter neutron drip line nuclei taking the ratio 
N/Z=2 and the region 10$\leq$N$\leq$60.   From Fig. 1 one sees the  
approximate neutron number at which spherical (or deformed) 
halo phenomena are expected to occur, and the reason why the halo may be 
observed.  The deformed halos around N$\approx$10 
(for example, $^{11}_{4}$Be$_7$) 
due to the almost degeneracy of the 2s$_{1/2}$ and 1d$_{5/2}$ levels and 
around N$\approx$24 (for example, $^{37}_{12}$Mg$_{25}$) due to 
the almost degeneracy of the 2p$_{3/2}$ and 1f$_{7/2}$ levels are 
already experimentally confirmed.  
Therefore, one may expect another spherical or deformed halo 
around N=51  
(for example, $^{77}_{26}$Fe$_{51}$ ).  
It is interesting to see in Fig. 1 that deformation may play 
a role almost whenever halo phenomena 
(with $s$ or $p$ components of neutrons) are expected.  
I note that due to the simplicity of the present one-body potential  
one can easily obtain a systematic view of the shell structure 
around neutron drip line, compared with sophisticated and elaborated 
microscopic models starting with effective two-body interactions.   

Second, I studied the neutron shell structure of lighter neutron-rich nuclei 
related to the r-process just above $^{78}_{28}$Ni$_{50}$, choosing N/Z=1.7 
and 40$\leq$N$\leq$90.  
If the proton numbers accept deformation, some even-even nuclei in the region 
of 60$\simleq$N$\simleq$70 in Fig. 2 may be prolately deformed among the nuclei 
in the major shell 50$\leq$N$\leq$82, 
following the mechanism similar to the case in which the stable
rare-earth nuclei with 90$\simleq$N$\simleq$112 are deformed among the nuclei 
in the major shell 82$\leq$N$\leq$126.   
The spherical shell structure 
around the Fermi level ($\varepsilon_{j} \approx -$4 MeV) for a given neutron 
number is compared with the shell structure of stable nuclei 
around the Fermi level 
($\varepsilon_{j} \approx -$8 MeV) for the same neutron number.   
Since the shell structure around the Fermi level of stable nuclei is 
well studied, one may use the idea that the information 
known in stable nuclei can be used in the analysis of r-process nuclei, 
when the neutron number is the same in the two cases.  
Taking the same neutron number around respective Fermi levels, 
the main difference of the shell structure of the r-process nuclei 
in this mass number region from that of stable nuclei is found to come
essentially from 
the difference between the Fermi energies, $\varepsilon_{j} \approx -4$ MeV 
versus $\varepsilon_{j} \approx -8$ MeV.   Namely, as the neutron number 
decreases (i.e. the potential becomes weaker), higher-$\ell$ orbits 
lose the binding energies more rapidly than lower-$\ell$ orbits.  
That means, 
within a given major shell the binding energies of higher-$\ell$ orbits 
around the Fermi energy of stable nuclei are larger than those of lower-$\ell$ 
orbits by an order of 1 MeV compared with the case around the Fermi energy 
of the r-process nuclei.  Though the order of 1 MeV is 
an appreciable change of 
the shell structure within a given major shell, 
due to the possible deformation in nuclei of Fig. 2 
and the finite temperature present 
in the r-process nuclei  
it is not trivial how much effect this kind of 
basic shell-structure difference in r-process nuclei from that 
in stable nuclei 
has on some observed quantities, before calculations
of some individual nuclei are quantitatively examined.   

Third, I presented the study of the shell structure  
for the potential 
provided by $^{70}_{26}$Fe$_{44}$ (N/Z=1.69), which is a representative 
example of so-called ''N=40 IoI'' nuclei.   It is experimentally indicated 
that even-even Cr isotopes with N=38-42 and even-even Fe isotopes with N=40-46 
are deformed.  
In the present model the possible deformation of Fe isotopes with 
40$<$N$\simleq$46 is expected, while deformation of the nucleus 
$^{66}_{26}$Fe$_{40}$ is not obtained because the Fermi level of 
the N=40 system in the region of moderately-deformed prolate shape 
never becomes
lower than that of spherical shape which is equal to $\varepsilon$(2p$_{1/2}$).  In this respect I note that though 
the observed E(2$_{1}^{+}$) = 574 keV in $^{66}$Fe 
is pretty small and may indicate deformation, the observed ratio of 
E(4$_{1}^{+}$)/E(2$_{1}^{+}$) is only 2.45.  
The very small ratio compared with 3.33 indicates that $^{66}$Fe is not  
a stably-deformed nucleus.    

Though in the literature it is wondered whether or not 
the N=40 IoI 
deformation extends to N=50 in the $_{24}$Cr and $_{26}$Fe isotopes, 
from Fig. 5 I conclude that the Fe and Cr isotopes may be 
prolately deformed only up to 
N$\approx$46 and the deformation contining up to N=50 is unlikely, because 
of a few MeV energy difference between the 2d$_{5/2}$ and 1g$_{9/2}$ orbits 
at $\beta$=0 in ''N=40 IoI'' nuclei in contrast to the almost degenerate 
2p$_{3/2}$ and 1f$_{7/2}$ orbits at $\beta$=0 in ''N=20 IoI'' nuclei.   The
consequence of this difference on the possible deformation in relevant nuclei 
is explained in detail in Sec. III. C.      

In the present work I have concentrated on the discussion of 
the fundamental shell 
structure of one-particle energy spectra and did not take into account 
the role of pair correlation.  The shell structure which I have discussed is 
anyway the first important quantity before including pair correlation.   
In the quantitative comparison with experimental data the inclusion of 
pair correlation may be necessary perhaps 
except for light halo nuclei such as those 
discussed in Sec. III. A.  
In some r-process nuclei discussed in Sec. III. B nonzero temperature 
together with deformation and 
pair correlation is further necessary to be included 
in the quantitative comparison with experimental information.  
In the region of ''N=20 IoI'' nuclei 
(for example, neutron-rich Mg isotopes) 
a model in line with the present article could well explain available
experimental data.      
Since the shell structure in the region of ''N=40 IoI'' is, in some part, 
similar to that in the region of ''N=20 IoI'' as discussed in Sec. III. C, 
it is interesting to see 
how far the result presented in Sec. III. C is in agreement with findings in 
experiments in the future.

\newpage

\noindent
{\bf\large Figure captions}\\
\begin{description}
\item[{\rm Figure 1 :}]
Calculated neutron one-particle energies in a spherically symmetric Woods-Saxon 
potential produced by nuclei with N/Z=2, which determines $V_{WS} = -$40 MeV.
The ratio N/Z=2 is an approximate ratio of the even-even core 
of the observed heaviest odd-N nuclei for a given even-Z in the lighter mass region.   
The Fermi levels of nuclei with respective neutron numbers lie around 
$\varepsilon_{j} = 0$.
\end{description}

\begin{description}
\item[{\rm Figure 2 :}]
Calculated neutron one-particle energies 
in a spherically-symmetric Woods-Saxon 
potential produced by nuclei with N/Z = 1.7, which determines $V_{WS} = -$42.4 
MeV.   
The ratio N/Z=1.7 is an approximate ratio in lighter nuclei related to 
the r-process just above $^{78}_{28}$Ni$_{50}$.  
The Fermi levels of nuclei with respective neutron numbers lie around 
$\varepsilon_{j} = -4$ MeV.
\end{description}

\begin{description}
\item[{\rm Figure 3 :}]
Calculated neutron one-particle energies 
in a spherically-symmetric Woods-Saxon 
potential produced by nuclei along the stability line.   The depth  of 
the Woods-Saxon potential is determined by the formula (\ref{eq:WS}) 
using the proton number given by respective neutron numbers along the stability 
line.   Thus, the depth of the potential changes 
as a function of neutron number. 
The Fermi levels of nuclei with respective neutron numbers lie around
$\varepsilon_{j} = -8$ MeV.  
\end{description}

\begin{description}
\item[{\rm Figure 4 :}]
Calculated proton one-particle energies in the potential produced by $^{70}$Fe 
as a function of quadrupole deformation $\beta$.  
Some proton numbers, which are obtained by filling all lower-lying levels, are 
indicated with open circles.  
For spherical shape ($\beta$ = 0) the quantum numbers ($n \ell$ j) of orbits 
are denoted 
except for 2p$_{1/2}$ at $\varepsilon_{j}$ = $-$9.16 MeV and 1f$_{5/2}$ at 
$\varepsilon_{j}$ = $-$10.24   MeV.   
\end{description}

\begin{description}
\item[{\rm Figure 5 :}]
Calculated neutron one-particle energies in the potential produced by $^{70}$Fe 
as a function of quadrupole deformation $\beta$.  
Some neutron numbers, which are obtained by filling all lower-lying levels, are 
indicated with open circles.
One-particle bound and resonant energies 
at $\beta = 0$ are $-$7.80, $-$6.52, $-$6.14, 
$-$2.99, $-$0.37, $-$0.25, +1.44, and +3.56 MeV 
for the 2p$_{3/2}$, 1f$_{5/2}$, 2p$_{1/2}$, 1g$_{9/2}$, 2d$_{5/2}$, 
3s$_{1/2}$, 2d$_{3/2}$, and 1g$_{7/2}$ levels, respectively.  
One-particle resonant levels ($\varepsilon_{\Omega} > 0$) for $\beta \neq 0$ 
are not plotted if they are not relevant to the present interest. 
\end{description}

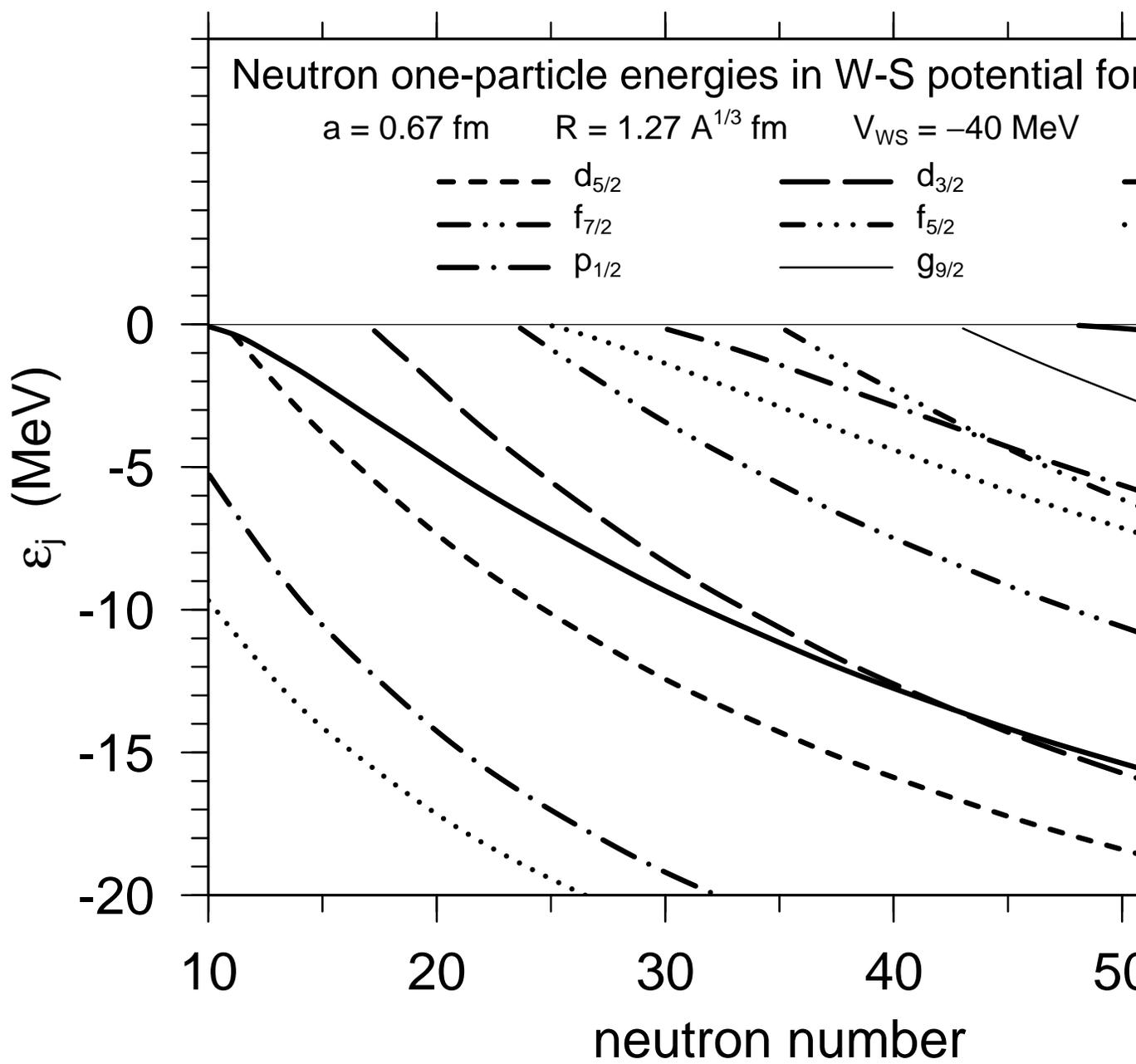
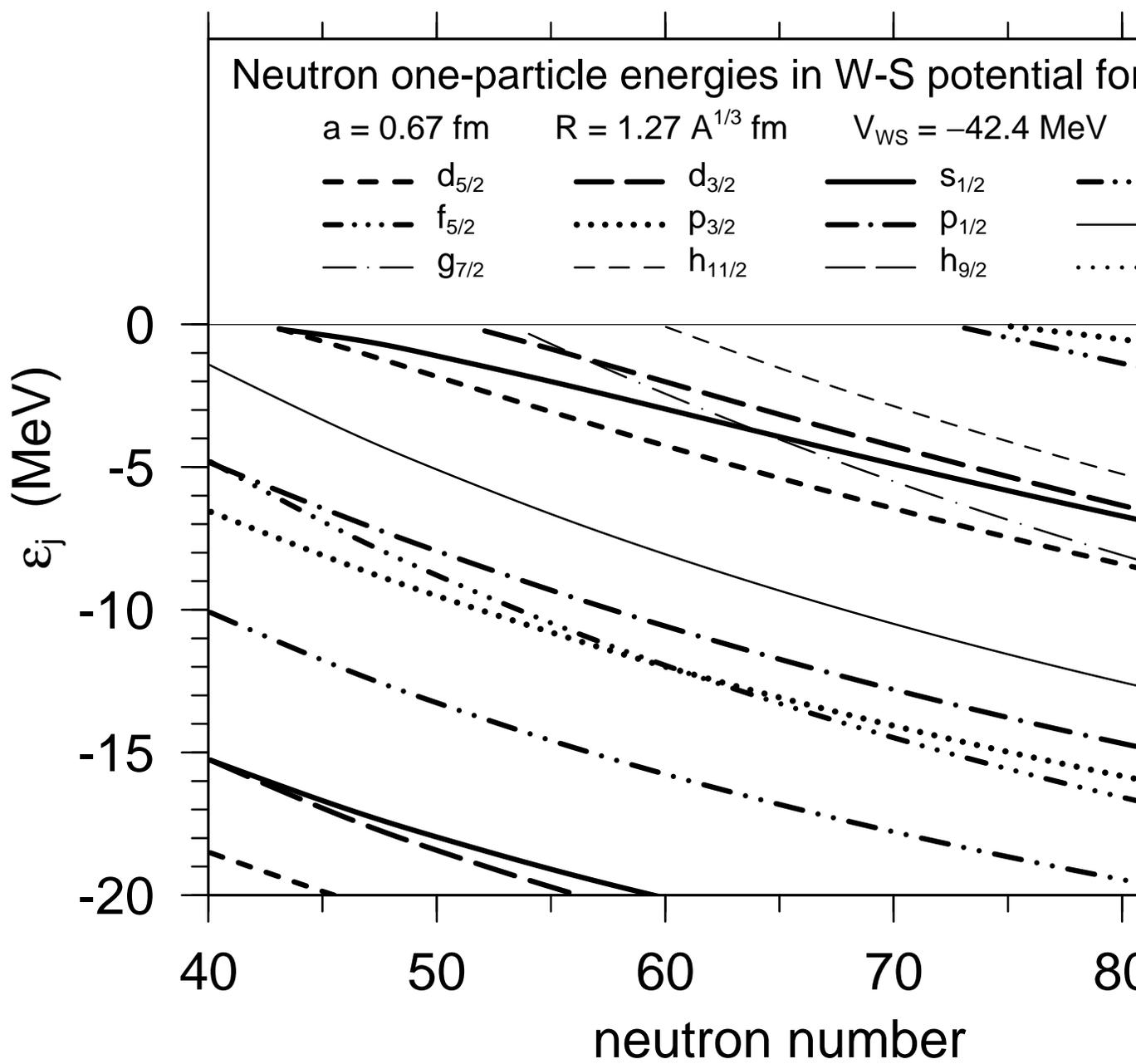
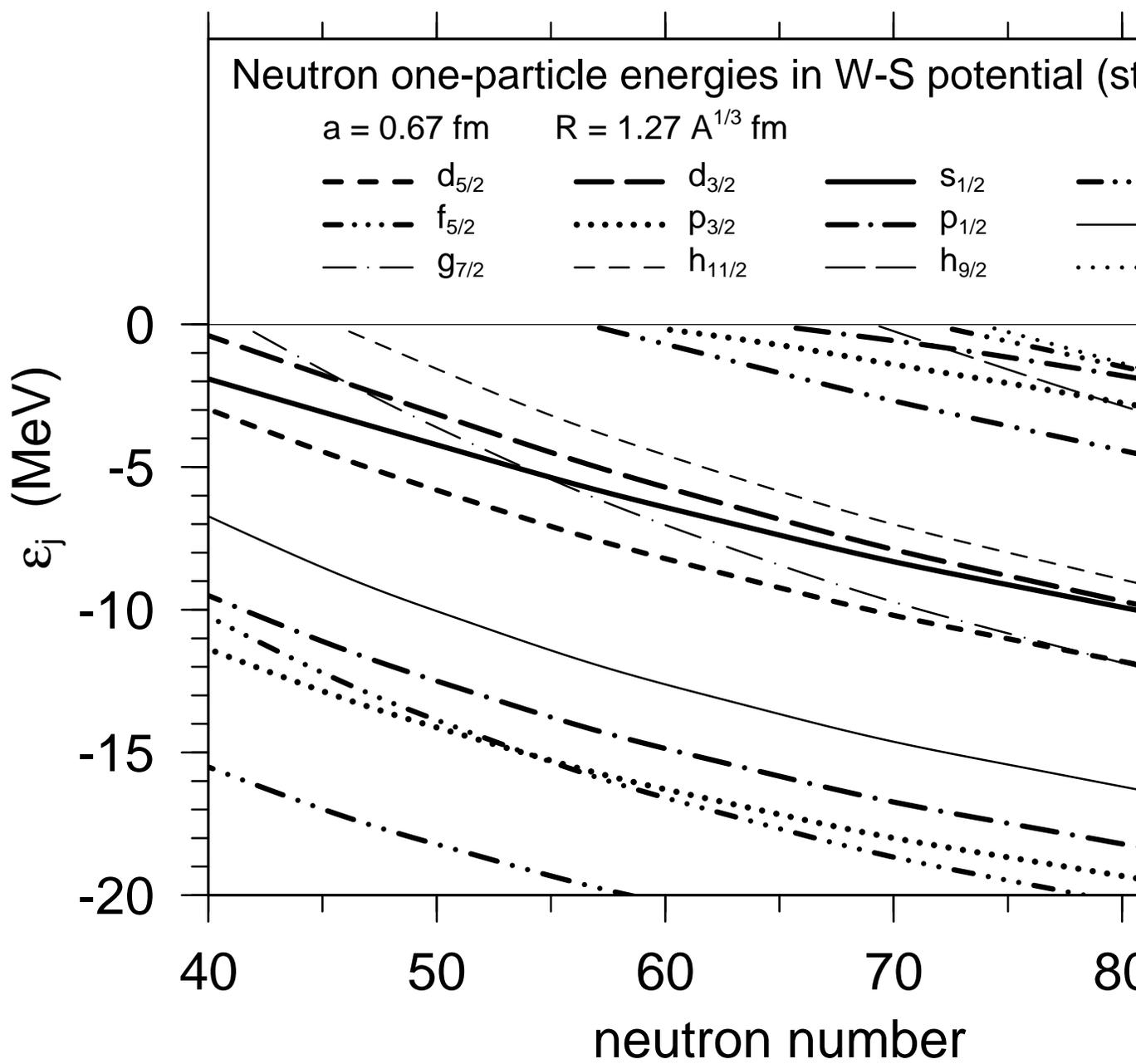
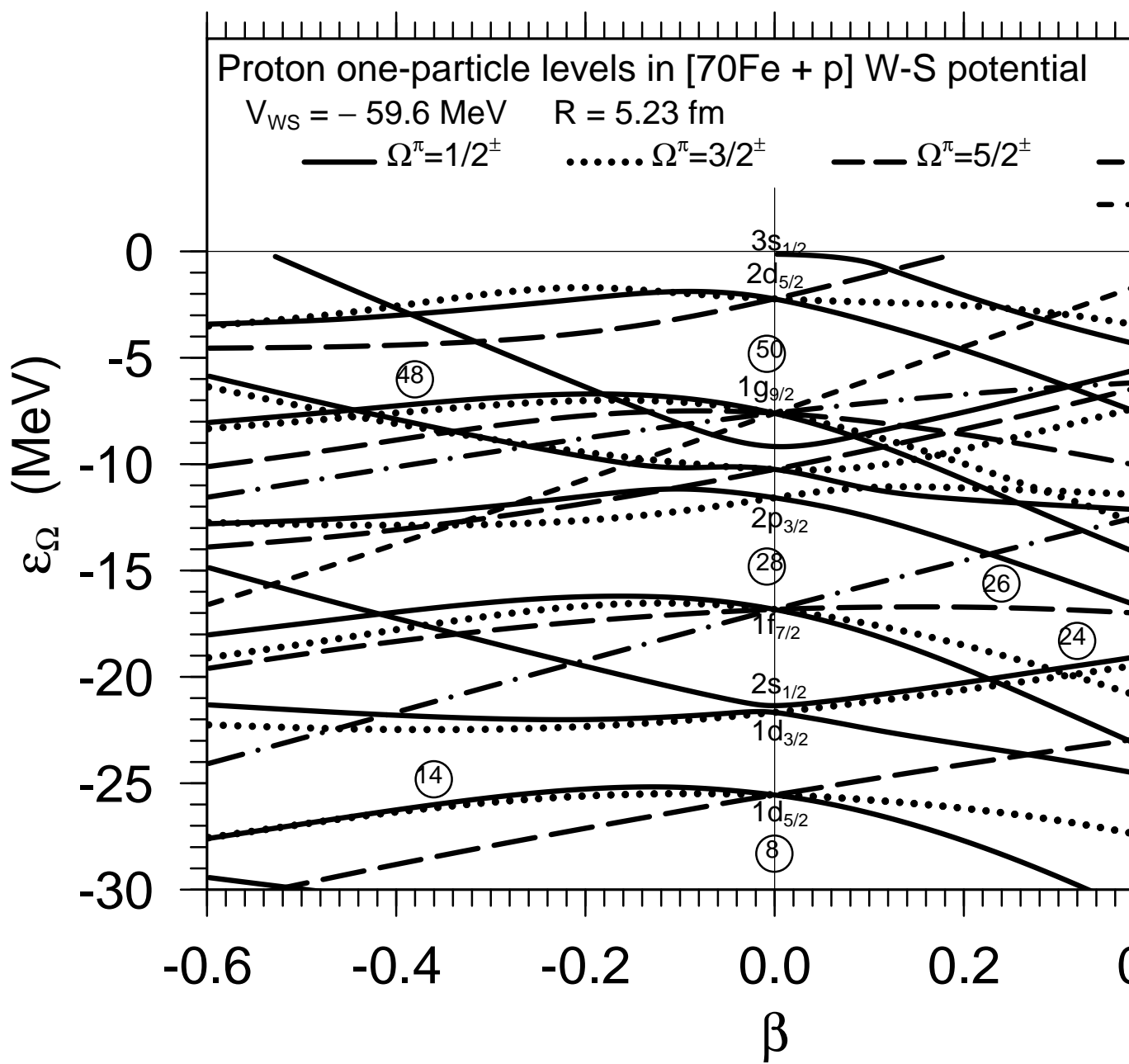
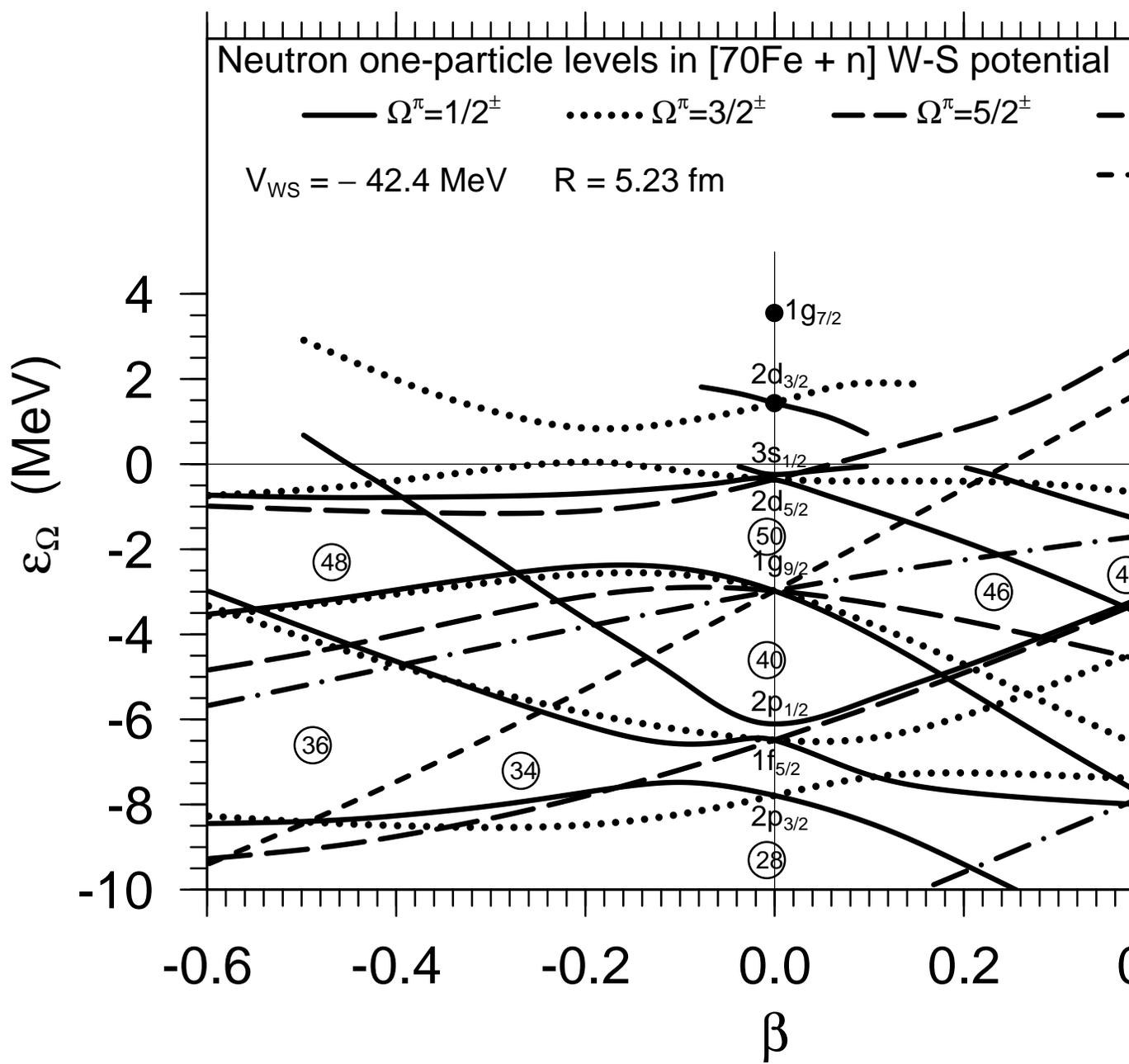
\end{document}